\documentclass[prb,rotate,showpacs]{revtex4}
\usepackage{graphicx}
\begin{document}
\author{Th. Theenhaus}
\author{M. Letz} \altaffiliation[Current address: ]{Schott Glas,
Research and Developement, Hattenbergstr. 10, 55014 Mainz, Germany}
\author{A.Latz}
\altaffiliation[Current address: ]{Institut f\"ur Physik, TU - Chemnitz,
D-09107 Chemnitz, Germany}   
\author{R. Schilling}
\affiliation{Institut f\"ur Physik, Johannes Gutenberg-Universit\"at, 
Staudingerweg~7, D-55099~Mainz, Germany}
\author{M. P. Allen}
\altaffiliation[Current address: ]{Centre for Scientific Computing, 
University of Warwick, Coventry CV4 7AL, UK}
\affiliation{H. H. Wills Physics Laboratory, University of Bristol,
Royal Fort, Tyndall Avenue, Bristol BS8 1TL,  UK}

\title{Dynamical precursor of nematic order in a dense fluid of hard
  ellipsoids of revolution.} 

\begin{abstract} 
We investigate hard ellipsoids of revolution in a parameter regime
where no long range nematic order is present but already  finite size
domains are formed which show orientational order. Domain formation
leads to a substantial slowing down of a collective
rotational mode 
which separates well from the usual microscopic frequency regime. A
dynamic coupling of this particular mode into all other modes provides
a general mechanism which explains an excess peak in spectra of
molecular fluids.  Using molecular dynamics simulation on up to 4096 particles
and on solving the molecular mode coupling equation we investigate
dynamic properties of the peak and prove its orientational origin.
\end{abstract}

\date{\today}
\maketitle   

If the temperature of a simple liquid is decreased or its density
increased local correlations tend to grow.  This tendency can be experimentally
detected as a growing peak of the static structure factor $S_q$ at a
wave vector $q=q_{max}= 2 \pi / \Delta r_{av}$, where $\Delta  r_{av}$
is an average nearest neighbor distance. The growth of the static
structure factor 
at $q_{max}$ causes the dynamic to slow down on local length scales,  a
phenomenon known in neutron scattering experiments as de Gennes
narrowing \cite{degennes59}. Hansen and Verlet \cite{hansen69}
observed in simulations, that a large class of simple liquids
crystallize, if the  peak of the static structure factor at
$q_{max}$  becomes larger than $2.80$. The growing peak may therefore
be interpreted as a precursor phenomenon of the first order
crystallization transition. But since the local dynamics is
slowing down, it can happen that the crystalline minimum of the free
energy is not
reached during the time scale of the experiment. The system becomes
supercooled and begins to exhibit the rich dynamics of supercooled
liquids \cite{goetze91,goetze92,schilling94,goetze99b}. Dense fluids of hard spheres are a good
experimental example, where this physics can be studied in great
detail.            

A simple example of a molecular liquid with translational and
rotational degrees of freedom is obtained by deforming the
hard spheres  towards hard ellipsoids of revolution. They can be
characterized by the diameter of the minor axes $d$ and the aspect
ratio $X_0$. This is the relation between the major and minor axes of
the ellipsoid. In the following discussion $d$ is always set to $d=1$,
i.e. length will be given in units of $d$, 
and  we will address  only prolate
systems with aspect ratios larger than one. 
However we expect qualitatively similar results for oblate systems,
replacing 
$X_0$ by $1/X_0$ \cite{letz99c}. 
If the aspect ratios of
the ellipsoids are very large, the theory of Onsager (originally for hard
sphero--cylinders) is valid \cite{onsager49}. This theory predicts a
weakly first  order phase transition from an isotropic
fluid to a nematic upon increasing the packing fraction $\varphi =
X_0  \rho \pi/6$, with  $\rho$ being the number density.  
Such a transition can be found from computer
simulations \cite{frenkel84} for aspect ratios
down to $X_0 \approx 2$. The weakly first order phase transition also
has a precursor 
in the static structure. The autocorrelation function $S_{22}^m(q
\rightarrow 0)$ of the
collective fluctuations of the second spherical harmonics, measuring
the quadrupolar order of the molecules is  strongly increasing, when the
transition is approached. 
This value of the correlation function is proportional to the optical
Kerr constant and related to the correlation function of the nematic
order parameter. Theoretically it is possible  to supercool
also the weakly first order phase transition of the nematic
instability \cite{letzschill99} although to our knowledge it has not
been observed yet. The increase of $S_{22}^m(q \rightarrow 0)$ when
increasing the 
packing fraction towards the isotropic to nematic transition leads,
analogous to 
the de Gennes narrowing phenomena, to a slowing down of the
corresponding dynamic orientational correlation function. 
In extreme cases such a collective mode can drive a glass transition
\cite{aksenov87,schilling97b,michel88,bostoen91}.  

Our molecular dynamics (MD) simulation is placed in the regime where an 
increase of the peaks of the static structure factor, and a formation
of nematic ordered  
domains, are both present. This has  been achieved by simulating  a
system of $N$  nearly hard ellipsoids with an  aspect ratio of $X_0 = 1.8$, 
where $N$ has been chosen to be 4096 and 512. 
Nearly hard ellipsoids are defined by a pair potential 
$v(1,2)=4\epsilon(s^{-12}-s^{-6}+\frac{1}{4})$ 
for $s<2^{1/6}$, $v(1,2)=0$ otherwise.  Here $s=(r(1,2)-\sigma(1,2)+d)/d$
is a scaled and shifted separation where the `diameter' $\sigma(1,2)$
is a standard approximation \cite{berne72} to the contact distance of
ellipsoids 1 and 2 at given relative orientation, and $r(1,2)$ is the 
center-center distance.  The potential is purely repulsive
and varies quite strongly over a short range of interparticle separation;
it has been used, for higher elongations $X_0$, in the study of nematic
liquid crystals \cite{mcdonald2000,akino2001}.  We choose an energy parameter
$\epsilon=1$, and molecular mass $M=1$. In each case
the system was equilibrated at a temperature
$k_BT=1$, and then constant-energy
molecular dynamics were carried out using timesteps in the
range $\delta t=$0.003--0.005; typical run lengths were $10^6$ timesteps.

From the simulation the 
molecular density--density correlation functions  $S_{ll'}^m(q,t)$ are
obtained.  
\begin{equation}
S_{ll'}^m(q,t) = \frac{1}{N} \langle \rho_{lm}^*(q,t) \,
\rho_{l'm}(q,0) \rangle 
\end{equation}
The indices $l,l',m$ are the indices of the spherical harmonics, the
brackets $\langle ... \rangle $ denote a statistical average and $
\rho_{lm}(q,t) = \sqrt{4 \pi} i^l \sum_{n=1}^N \exp\{i
\vec{q}\vec{x}_n(t)\} Y_{lm}(\Omega_n(t))$ is the time dependent
tensorial 
density fluctuation, where $ \vec{x}_n(t)$ and $\Omega_n(t)$ are position and
orientation of the molecule $n$, respectively. 
We also have chosen a specific
coordinate system, the q-frame, where the $z$-axis
points along the 
wave-vector ${\bf q}$. In this particular coordinate system all
correlation functions  depend on $l$ and $l'$ and only on a single
helicity index $m$ and on the 
absolute value $q = | {\bf q}|$.
The static structure factors are  the equal time correlation function and reads
\begin{equation}
S_{ll'}^m(q) = \lim_{t \rightarrow 0} S_{ll'}^m(q,t) 
\end{equation}
For all theoretical considerations shown in this work we truncate the
correlation functions at $l_{max}=m_{max}=2$. 

\begin{figure}
\centerline{\rotatebox{0}{\resizebox{8cm}{!}{\includegraphics{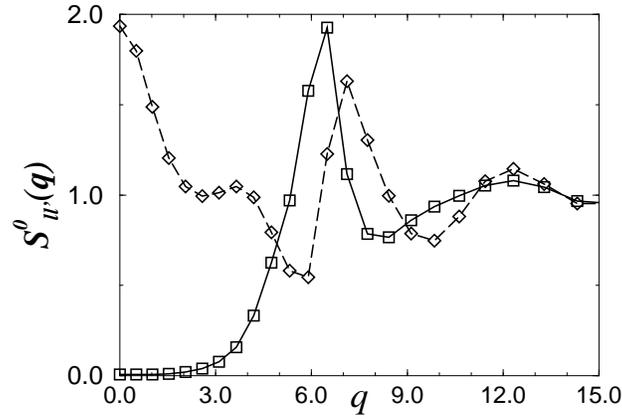}}}}
\caption{For an aspect ratio of $X_0=1.8$, from a MD simulation of
4096 hard ellipsoids of revolution the $S_{00}^0(q)$ component (solid
line) and the $S_{22}^0(q)$ (dashed line) components of the static
structure are  shown. The packing fraction was $\varphi=0.575$  
and is right at the glass transition from MMCT. 
}
\label{fig:1} 
\end{figure}

The MD simulation is first used to generate the static structure factors.
In Fig. \ref{fig:1} for a packing fraction of $\varphi = 0.575$, 
below any
transition the $S_{00}^0(q)$ correlator is plotted together with the  
$S_{22}^0(q)$ correlator. The dense liquid manifests itself in the low
compressibility $\kappa = \kappa_0 \lim_{q\rightarrow 0} \;
S_{00}^0(q)$. $ \kappa_0$ is the compressibility of an ideal
gas. Important there is the  well pronounced peak at $q=q_{max} =
6.5$, discussed above. 
This peak is the static manifestation of the cage formed by the
nearest neighbors which is about to trap the particles. At higher
densities it drives a glass transition. In the $S_{22}^0(q)$
correlator 
shown in the same plot
an additional peak occurs at  $q =0$. The reader should also note that
the peak at $q=0$, which is a precursor of nematic order is
accompanied by a second peak at $q \sim q_{max}$ and a third peak at
about the same position as the third peak in $S_{00}^0(q)$.    
The half width of the peak at $q=0$ 
is approximately unity $\Delta q \approx 1$ and indicates the
formation of orientationally ordered domains with a diameter of $D= 2
\pi / \Delta q \approx 6 $. Such a domain formation is
still far away from a true nematic transition. Yet, the $S_{22}^0(q)$
correlator reaches already a value of $1.9$ at $q=0$.  The static structure
factors are used as input for the equations of the molecular mode
coupling theory for the dynamic structure factors of supercooled molecular
liquids \cite{schilling97}. Some details of this theory and an 
application to dipolar hard spheres can be found in
\cite{theenhaus01}. For the comparison with the simulation it is
important to remember, that the mode coupling theory of supercooled
liquids overestimates the slowing down of structural
relaxations upon decreasing the temperature or increasing the
density \cite{pusey87,vanmegen91,barrat90,nauroth97,winkler00}. 
To allow for comparison with the simulations we have to
rescale the packing fractions for the MMCT such, that the time scale
of structural relaxations of the MMCT corresponds to the time scale of
the structural relaxation in simulations.

The static structure factors in dense liquids determine directly the 
microscopic frequencies measured in experiments of the respective
dynamic structure factors. Neglecting the coupling to heat
fluctuations in the liquid,  the matrix of microscopic
frequencies  is given by

\begin{equation}
\label{eq:omega}
(\underline{\underline{\Omega}}(q)^2)_{ll'}^m = 
\left ( q^2 \frac{k_BT}{M} + l(l+1) \frac{k_BT}{I} \right )
(\underline{\underline{S}}^{-1}(q))_{ll'}^m
\end{equation}
where $I$ is the moment of inertia of the ellipsoid with respect
to a rotation on one of the minor axes and $M$ the mass. For $l=l'=0$
Eq. (\ref{eq:omega}) describes the longitudinal (isothermal) phonon
frequency with 
its linear dispersion for small $q$-values; $\Omega_{00}^0(q) = c \,
q$ with the isothermal sound velocity of the liquid $c$. For $l,l' \ne
0$ and $q \to 0$ the static and dynamic structure factors become diagonal in
$l,l'$ and independent of $m$ for short range interaction in general
and the hard ellipsoids in particular. Therefore the  microscopic
frequencies in (\ref{eq:omega})  reduce for $l,l' \ne 0$ and $q=0$ to 
\begin{equation}
\label{eq:omega_opt}
\{\underline{\underline{\Omega}}(q)^2\}_{ll'}^m = 
\left (l(l+1) \frac{k_BT}{I} \right ) \frac{1}{\{
\underline{\underline{S}}(q =0)\}_{ll}^m} \delta_{ll'} 
\end{equation}

For the following it is important to note, that large values of the
static structure factors lead to small microscopic frequencies and the
corresponding slow modes have a strong intensity proportional to the
corresponding static structure factor.  This
is the origin of  the de Gennes narrowing phenomenon in the center of
mass motion at wave vectors around $q =q_{max}$. The increase of the
static structure factor $S_{22}^m(q=0)$ shown in Fig.\ref{fig:1}, 
will similarly
yield, due to the relation (\ref{eq:omega_opt}), 
a slowing down of the corresponding optical $l=2$ mode. 

As soon as the structural relaxation becomes much slower than the
microscopic motion, the microscopic frequencies become renormalized
since the system is in a glassy state on the time scale of the
microscopic motion. For molecular liquids also a splitting of the
microscopic frequencies occur in addition to the renormalization due
to rotation translation  coupling   \cite{theenhaus01,theenhaus01b}.  But the
slowing down of the rotational mode due to the occurrence of locally
ordered domains is preserved also under renormalization
\cite{theenhaus01b}.

Nonlinear memory effects will couple all these modes. 
Therefore the described
slow but intense modes will influence the spectra at different
wave vectors and different rotation numbers $l$ in the following way. 
The center of
mass motion on a spatial scale of about  $\Delta r \approx 1/q_{max}$
will be the primary reason for a
trapping of a particle in the cage of its nearest neighbor which leads
to the bifurcation scenario described by MCT
\cite{bengtzelius84}. Indirectly, the precursor of nematic order seen
in the static correlation function $S_{22}^m(q)$ for $q=0$ can support the
tendency towards glass formation via the mentioned induced local
spatial order, which is manifested as a peak in  $S_{22}^0(q)$ at $q
\sim q_{max}$.   
The fact that
there is a mode of strong oscillator strength and low frequency,
$S_{22}^m(q \approx 0,t)$, can  lead to a dynamic signature of this
mode in other modes. It can be recognized by varying the inertia
moment $I$. If and only if an excitation in a density spectrum at
some wave vector $q
\ne 0$ is due to the  rotational mode at
$q=0$, its position will shift according to Eq. (\ref{eq:omega_opt})
proportional to $1/\sqrt{I}$.   
The moment of inertia can be manipulated independently
from the mass of the ellipsoids in the MD simulation and the
numerical solution of the MMCT equations.  Even extremely high values can be
generated which can not be realized in a real systems. 

\begin{figure}
\centerline{\rotatebox{0}{\resizebox{8cm}{!}{\includegraphics{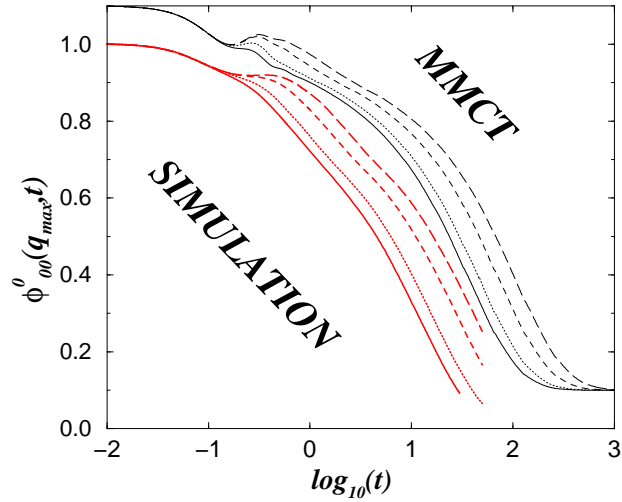}}}}
\caption{For four different values of the moment of inertia
$I_0$ (solid lines), $2.5 \times I_0$ (dotted lines) , $10
\times I_0$
dashed lines) , $25 \times I_0$ (long dashed lines) the time
dependent relaxation of $\phi_{00}^0(q=q_{max},t)=
S_{00}^0(q=q_{max},t)/S_{00}^0(q=q_{max},t=0)$ has been plotted. 
The upper curves are shifted by $0.1$ vertically and
result from a numerical solution of the MMCT
equations at $\varphi=0.56 $ and the lower curves are the result of a
MD simulation with 4096 particles at $\varphi=0.66$.
}
\label{fig:2} 
\end{figure}

In Fig. \ref{fig:2} the time dependent center of mass
correlation function  $l,l' =0$ at $q=q_{max}$ has been plotted. 
For the moment of inertia of an ellipsoid with homogeneous mass
distribution  $I_0 = (1 +
X_0^2)/20$, we have fixed two densities ($\varphi_{Sim}=0.66$,
$\varphi_{MMCT}=0.56$)  which lead to
approximately the same time scale for the structural relaxation in the
simulation and the numerical solution of the MMCT, respectively. Then,
the moment of inertia was increased to $2.5$, $10$  and  $25$ times
$I_0$. The simulation and the theoretical calculation exhibit a
shift to longer relaxation times, which are of the same order in both
cases. In addition to a shoulder at longer times a plateau develops
when the inertia moment is increased. 
This demonstrates clearly the considerable qualitative
influence of the rotational 
motion on the center of mass motion.  To study this effect and its
dependence on wave vectors more
quantitatively  it is instructive to calculate 
the susceptibility spectrum  ${\chi''}_{00}^0(q,\omega) = \omega
{\phi''}_{00}^0(q,\omega)$.  

\begin{figure}
\centerline{\includegraphics[width=170mm,angle=0]{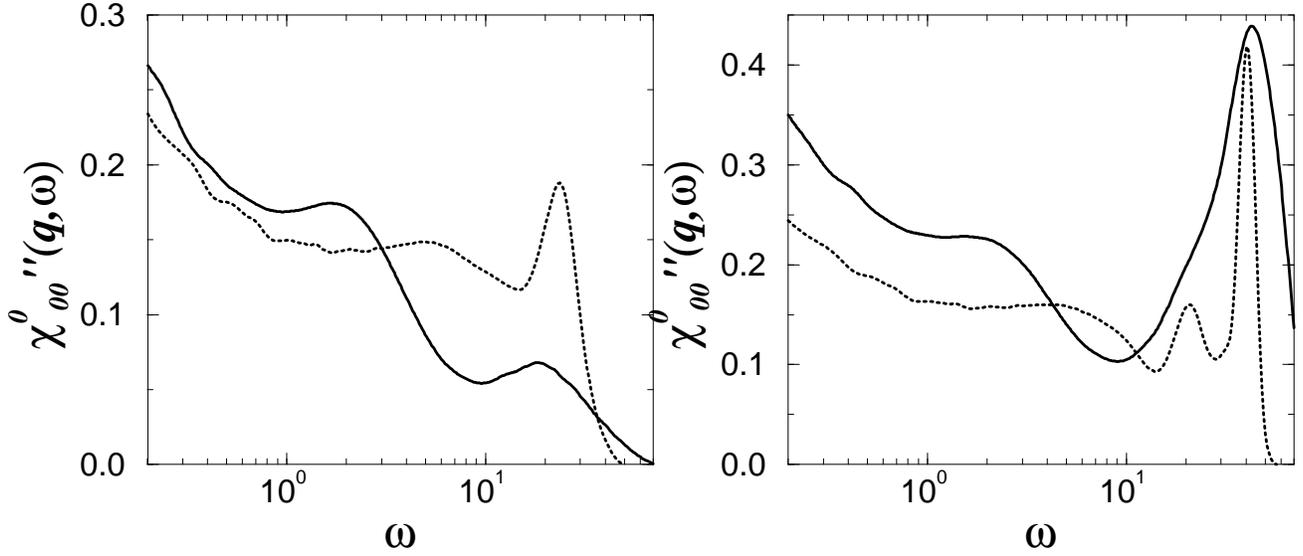}}
\caption{The susceptibility spectrum ${\chi''}_{00}^0(q,\omega)$ from
  simulation at packing fraction $\varphi =0.66$ (solid line) and MMCT
  at packing fraction $\varphi=0.56$ (dashed line) for two
  different wave vectors a)  $q_1\simeq 6.5$ and b) $q_2\simeq 2.71$.}   
\label{fig:3} 
\end{figure}

In Fig. \ref{fig:3} the susceptibility spectrum
${\chi''}_{00}^0(q,\omega)$ of the MMCT
result and the simulation are shown for two different
wave vectors. Note that there are no adjustable parameters. For
both wave vectors the wave vector dependent position of the high
frequency excitations agree 
very well. The splitting of the high frequency band, seen for  $q= 2.71$
in the MMCT calculation and mentioned above is smeared out in the
simulation. This may  be understood as follows.  Since we did not want to
bias our calculation with a theory for the microscopic bare damping,
we assumed a free oscillatory microscopic motion with frequencies
given by (\ref{eq:omega}). The effect of smaller damping can also be
seen in the time dependent functions (see Fig. \ref{fig:2}) as a tendency
towards oscillation in the plateau regime of the MMCT results. 
 By switching on a phenomenological
microscopic constant damping term $\nu_{ll}^m$ in the MMCT equations 
the split microscopic frequency  can be forced to merge in one
single excitation as in the simulation. In addition to the microscopic
excitation an additional almost wave vector independent hump can be
seen in the MD - and the MMCT result at
about one decade  below
the microscopic frequency band, before the $\alpha$ - relaxation peak
starts to grow at smaller frequencies.  This hump is due to the additional
plateau in Fig. \ref{fig:2}. 
 But the positions of the peak do not agree very
well quantitatively. The reason for this is not
completely clear. The effect can partially be accounted for by the 
missing microscopic damping in the MMCT calculation, since the
first minimum below the microscopic band in the simulation is already
shifted  to smaller frequencies because of damping. In addition, since
the memory kernels in the MCT - equations, which generate the coupling
between translational and rotational modes and therefore the hump, are
approximate, only, we can in general not expect a quantitative
agreement of both positions.

\begin{figure}
\centerline{\rotatebox{0}{\resizebox{8cm}{!}{\includegraphics{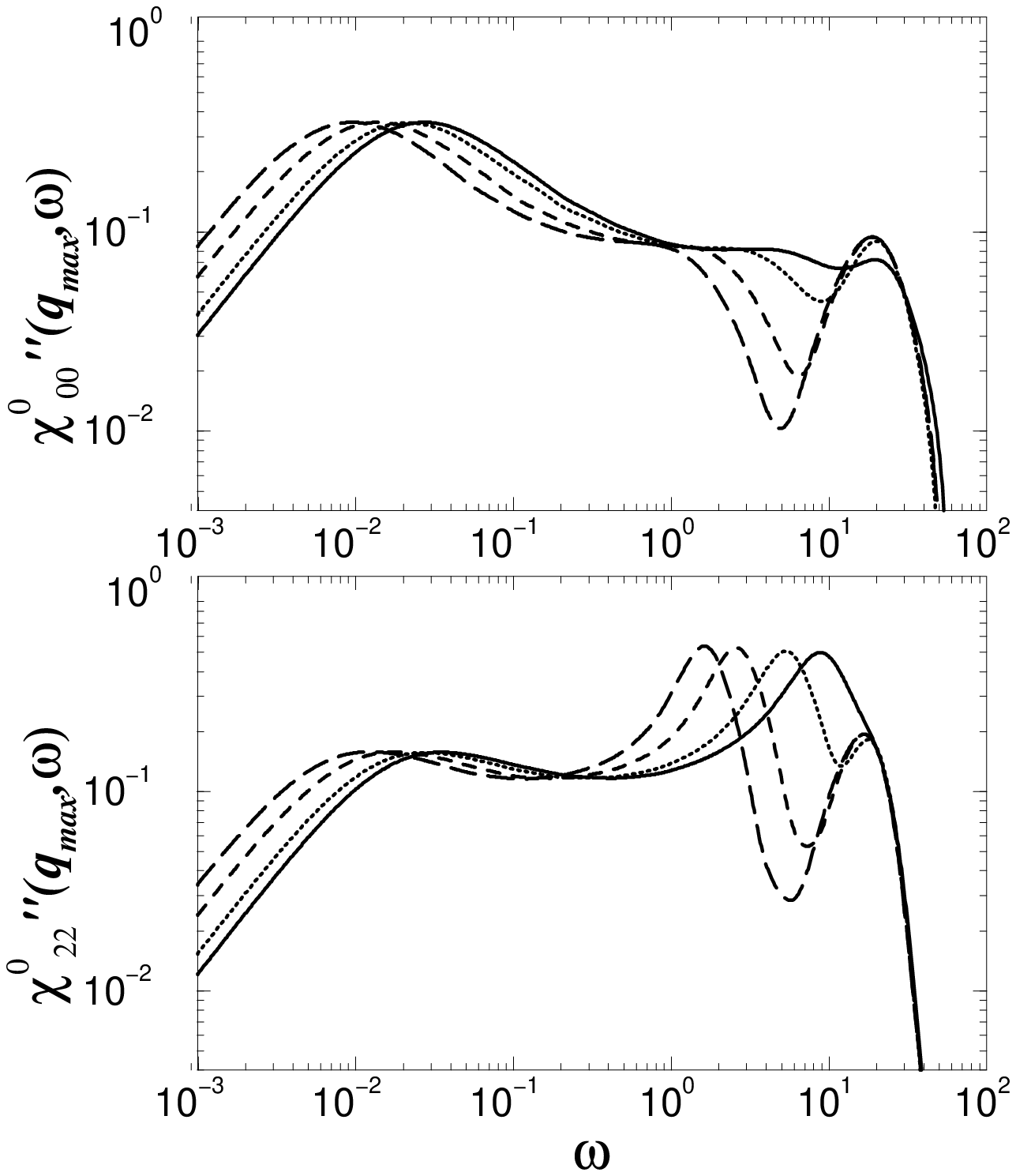}}}}
\caption{Susceptibility spectra ${\chi''}_{ll'}^m(q_{max},\omega)$ at the
  wave vector $q_{max}\simeq 6.5$ from the solution of the MMCT  at a packing
  fraction $\varphi=0.56$. The plots for four values of the moment of
  inertia $I_0$ (solid 
lines), $2.5 \times I_0$ (dotted lines), $10 \times I_0$ 
(dashed lines), $25 \times I_0$ (long dashed lines) are shown.}
\label{fig:4} 
\end{figure}

\begin{figure}
\centerline{\rotatebox{0}{\resizebox{8cm}{!}{\includegraphics{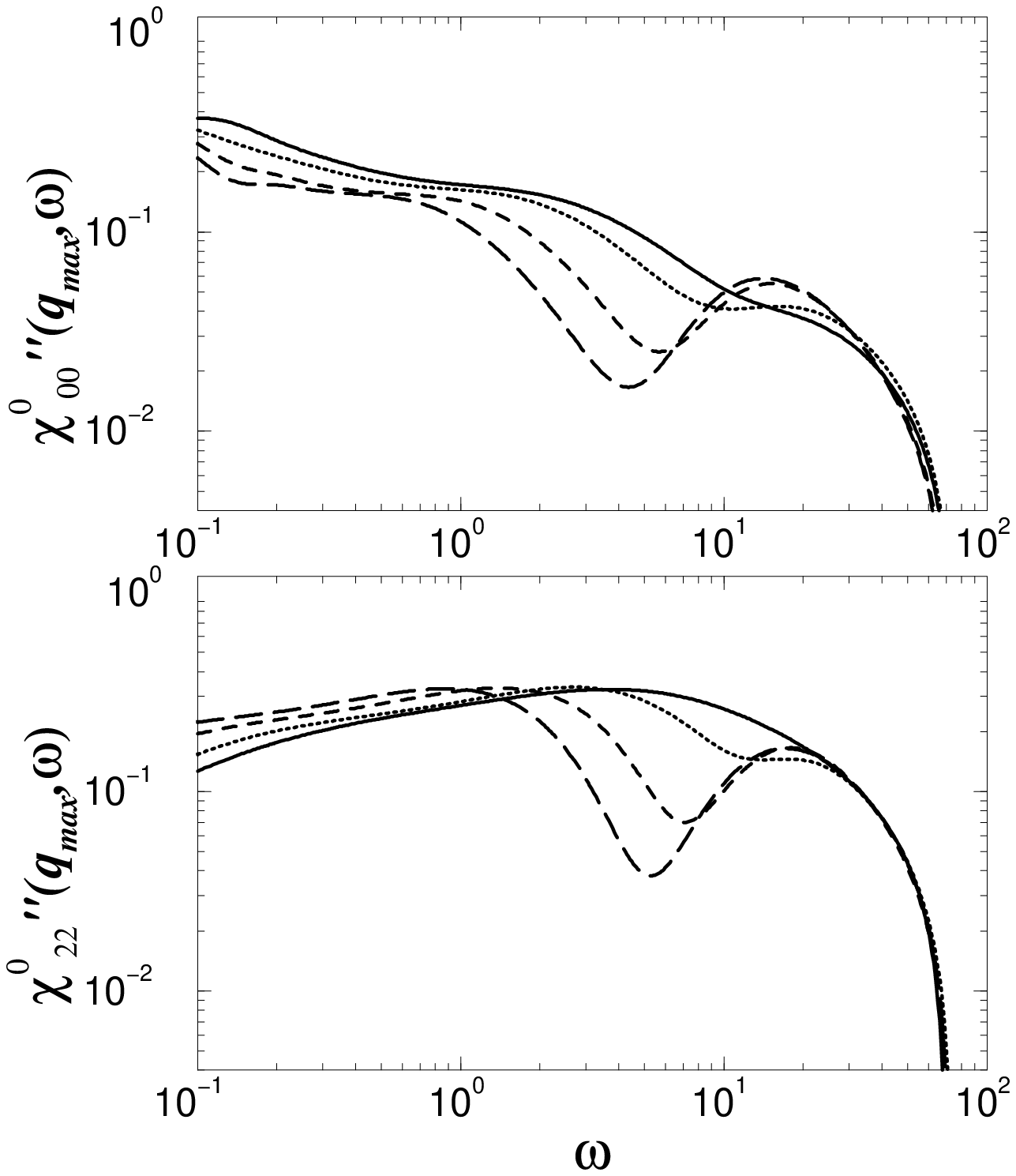}}}}
\caption{Susceptibility spectra ${\chi''}_{ll'}^m(q_{max},\omega)$ at the
  Debye wave vector $q_{max}\simeq 6.5$ from the MD simulation at a packing
  fraction $\varphi=0.67$. The plots for four values of the moment of
inertia $I_0$ (solid
lines), $2.5 \times I_0$ (dotted lines), $10 \times I_0$ 
(dashed lines), $25 \times I_0$ (long dashed lines) are shown.}
\label{fig:5} 
\end{figure}

\begin{figure}
\centerline{\rotatebox{0}{\resizebox{8cm}{!}{\includegraphics{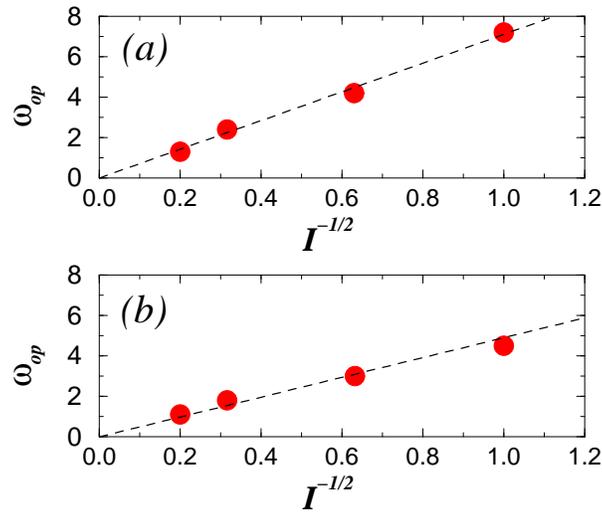}}}}
\caption{The average position of the anomaly in the five spectra
${\chi_{ll'}^m}''(q,\omega)$ is  plotted against
  $1/\sqrt{I}$. Shown are a) the results for the MMCT and b) the
results for the simulation.}
\label{fig:6} 
\end{figure}

The variation of the peak position of the low lying excitation in the
MMCT - result with
changing the  moment of inertia can be
seen in Fig. \ref{fig:4}. The susceptibility spectra
${\chi''}_{ll'}^m(q_{max},\omega)$ at the 
wave vector $q_{max} = 6.5 \approx$ exhibit the  anomaly between the
structural relaxation peak and the microscopic excitations for all
allowed combinations of $l,l',m$,  either as an additional  
peak or an additional  shoulder.  
 Especially in the $l=l'=2$
susceptibility spectra a well pronounced peak can be identified, which
shifts to lower 
frequencies for increasing  moment of inertia $I$ and is therefore
influenced by the orientational motion.  The $l=l'=0$
spectrum demonstrates that this orientation anomaly is hybridized with
the lowest of the split high frequency excitations for  $I =
I_0$ and turns into a well defined shoulder for larger inertia
moments. These features are
present in the spectrum of the MMCT calculations and the MD
- simulations for all susceptibilities $\chi_{ll'}^m$ with $(ll'm) =
(000),(200),(220), (221)$ and $(222)$. We show some of them in  
Fig. \ref{fig:5}.  
Since the MMCT calculations can be extended to longer
times compared to the simulations, the quality of the spectra are
better for the MMCT. But is is still possible to identify in
Fig. \ref{fig:5} for
all  moments of inertia the position of the anomaly.
To get better
statistics, we average the value of  the position
$\omega_{op}$ over five spectra $(ll'm) =
(000),(200),(220), (221)$ and $(222)$  and plot the results against
$1/\sqrt{I}$  in Fig. \ref{fig:6}. A well defined linear law
through the origin is found. This proves  our assertion that the
orientational anomaly in the $q=q_{max}$ spectra is due to the $q=0$
anomaly in $S_{22}^m(q=0)$ as  explained above. 

\begin{figure}
\centerline{\rotatebox{0}{\resizebox{8cm}{!}{\includegraphics{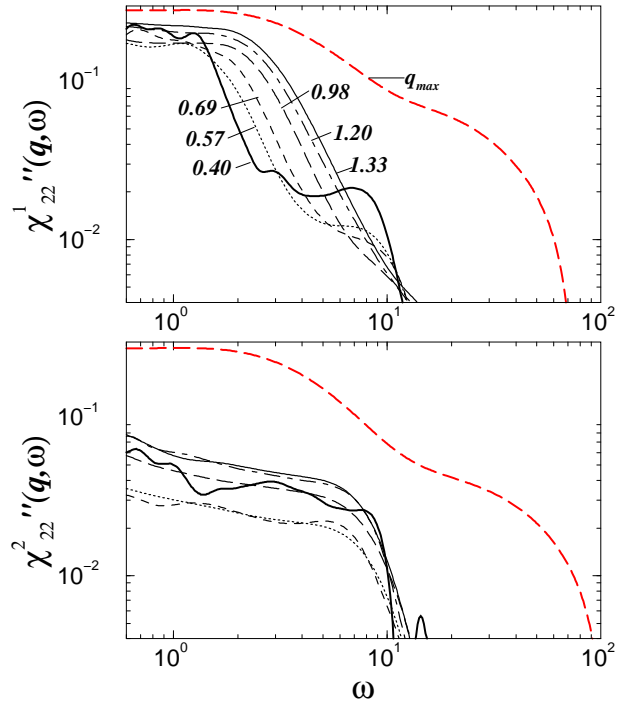}}}}
\caption{Susceptibility spectra ${\chi''}_{22}^1(q,\omega)$  and
${\chi''}_{22}^2(q,\omega)$ from the simulation at small wave vectors
$0.4 \leq q \leq 1.33$ and at $q=q_{max}$ and $\varphi = 0.69$.}
\label{fig:7} 
\end{figure}

The ${\chi''}_{22}^1(q,\omega)$ susceptibility  of the simulation shows
an additional 
excitation compared to $m=0$ or $m=2$, which is not present in the
version of MMCT used in this paper. This excitation is strongly
wave vector dependent (see Fig. \ref{fig:7}) and turns into a shoulder
for the three smallest wave vectors.  
It therefore does not have the same origin as
the orientation anomaly, which is almost wave vector independent.  
A possible explanation could be  a
coupling to transverse phonon modes. Such a coupling to transverse
phonon modes is expected from a more general version of the MMCT
\cite{latz00}. Unfortunately the allowed wave vector range of the
simulation seems to be too small to test for the linear dispersion
$\omega_T= c_T q$  of the expected coupling to transverse phonons  in the
spectrum ${\chi''}_{22}^1(q,\omega)$. 

In conclusion  we find for the spectra ${\chi_{00}^0}''(q,\omega)$ and
${\chi_{22}^0}''(q,\omega)$ a reasonably good agreement between the
MMCT for hard
ellipsoids and simulations of this system. We can identify a dynamic
anomaly which appears about one decade below the microscopic
excitations. This anomaly is due to the medium range nematic order
which can be identified by  the $q=0$ value of the static structure
factor $S_{22}^m(q=0)$. In \cite{theenhaus01} an analogous anomaly for
dipolar hard spheres due to medium range ferroelectric order could
be identified. There it was argued that the anomaly behaves very
much like the well known Bose-peak phenomenon. Similar arguments can be
applied to the anomaly caused by the medium ranged nematic order,
found here for hard ellipsoids.

\begin{acknowledgments}
We gratefully acknowledge financial support by
the "Sonderforschungsbereich 262" (Deutsche
Forschungsgemeinschaft). 
Computing facilities for the molecular dynamics simulations
were supported by the Engineering and Physical Sciences Research Council.
MPA gratefully acknowledges the
support of the Alexander von Humboldt Foundation. 
\end{acknowledgments}


\newpage

\end{document}